\RequirePackage{fix-cm}
\documentclass[smallcondensed]{svjour3}     % onecolumn (ditto)
\smartqed  % flush right qed marks, e.g. at end of proof
\usepackage{graphicx}
\usepackage{subcaption}

\usepackage[misc]{ifsym}
\usepackage{mathptmx}      % use Times fonts if available on your TeX system
\usepackage{hyperref}

\usepackage{amssymb}

\usepackage{float}

\usepackage{booktabs}
\usepackage{multirow}
\usepackage{makecell}
\usepackage{threeparttable}
\usepackage{color}

\usepackage[section]{placeins}

\makeatletter
\newif\if@restonecol
\makeatother

\usepackage[linesnumbered,ruled,vlined]{algorithm2e}%[ruled,vlined]{
\usepackage{algpseudocode}
\usepackage{amsmath}
\usepackage{cite}
\hypersetup{
    colorlinks=true,
    linkcolor=blue,
    filecolor=blue,      
    urlcolor=blue,
    citecolor=blue,
}
\usepackage[sort&compress,numbers]{natbib}
\begin{document}

\title{Dual Perceptual Loss for Single Image Super-Resolution Using ESRGAN%\thanks{Grants or other notes
}


\author{Jie Song\textsuperscript{1}           \and
        Huawei Yi\textsuperscript{1}          \and
		Wenqian Xu\textsuperscript{1}   \and
        Xiaohui Li\textsuperscript{1}   \and
		Bo Li\textsuperscript{1}    \and
		Yuanyuan Liu\textsuperscript{2} 
}

\authorrunning{J. Song et al.} % if too long for running head

\institute{
\Letter \,\,\, Huawei Yi \at
           \indent \,\,\,\,\,\,\,\,\,  yihuawei@126.com               %  \\
			\and
\indent \,\,\,\,\,\,\,\,\, Jie Song \at
           \indent \,\,\,\,\,\,\,\,\,  jiesong666666@163.com           %  \\
			\and
\indent \,\,\,\,\,\,\,\,\, Wenqian Xu \at
           \indent \,\,\,\,\,\,\,\,\,  x13275271957@163.com   
		    \and
\indent \,\,\,\,\,\,\,\,\, Xiaohui Li \at
           \indent \,\,\,\,\,\,\,\,\,  lhxlxh@163.com 
           \and
\indent \,\,\,\,\,\,\,\,\, Bo Li  \at
           \indent \,\,\,\,\,\,\,\,\,  leeboo@yeah.net  
			\and
\indent \,\,\,\,\,\,\,\,\, Yuanyuan Liu  \at
           \indent \,\,\,\,\,\,\,\,\,  liuyuanyuan@s.dlu.edu.cn  \\
			\and
			\textsuperscript{1}\,\,\,\,\,\,\, School of Electronics and Information Engineering, \at
			\indent \,\,\,\,\,\,\,\,\,\, Liaoning University of Technology, Jinzhou, 121001, China 
			\and
			\textsuperscript{2}\,\,\,\,\,\,\, Key Laboratory of Advanced Design and Intelligent Computing, Ministry of Education, \at
			\indent \,\,\,\,\,\,\,\,\,\, Dalian University, Dalian, 116622, China
}

\date{Received: date / Accepted: date}

\maketitle

\begin{abstract}
%Insert your abstract here. Include keywords, PACS and mathematical
%subject classification numbers as needed.
% 内容已缩减
The proposal of perceptual loss solves the problem that per-pixel difference loss function causes the reconstructed image to be overly-smooth, which acquires a significant progress in the field of single image super-resolution reconstruction. Furthermore, the generative adversarial networks (GAN) is applied to the super-resolution field, which effectively improves the visual quality of the reconstructed image. However, under the condtion of high upscaling factors, the excessive abnormal reasoning of the network produces some distorted structures, so that there is a certain deviation between the reconstructed image and the ground-truth image. In order to fundamentally improve the quality of reconstructed images, this paper proposes a effective method called Dual Perceptual Loss (DP Loss), which is used to replace the original perceptual loss to solve the problem of single image super-resolution reconstruction. Due to the complementary property between the VGG features and the ResNet features, the proposed DP Loss considers the advantages of learning two features simultaneously, which significantly improves the reconstruction effect of images. The qualitative and quantitative analysis on benchmark datasets demonstrates the superiority of our proposed method over state-of-the-art super-resolution methods.
%\subclass{MSC code1 \and MSC code2 \and more}
\end{abstract}

\section{Introduction}
\label{sec:1}
%Your text comes here. Separate text sections with
% 记得改引用
A low-resolution (LR) image is reconstructed as a high-resolution (HR) image, which is called single image super-resolution (SISR) reconstruction. It is an extremely challenging task. Under the condition of high upscaling factors, the lack of LR information leads to a series of uncertain factors, which will make the reconstructed image have a certain deviation from the ground-truth (GT) image. The main challenge of this paper is to make the reconstructed image as close as possible to the GT image, which is of great significance to the fields such as medicine \citep{1,2}, remote sensing \citep{3}, monitoring \citep{4} and so on.
\par
As the field of super-resolution (SR) reconstruction attracted more and more attention, many excellent algorithms have been proposed. According to the existing studies, the algorithms can be divided into three categories: interpolation-based methods, model-based methods and learning-based methods. The interpolation-based methods (e.g., bicubic interpolation \citep{5} and the nearest neighbor interpolation \citep{6}) have low complexity and high efficiency. However, since the expanded pixels are calculated by the neighborhood pixel, these methods don't make any reasoning for the texture details, which makes the image smooth on the whole. The model-based method can generate high-quality images efficiently, but this method is dependent on prior images relatively. When there is a certain deviation between the image to be reconstructed and the prior image, the visual quality of the reconstructed image will decrease rapidly. However, most SR methods are based on the learning, including neighbor embedding \citep{7}, sparse coding \citep{8,9,10,11} and random forests \citep{12}. With the popularization and development of deep neural network, the field of SR reconstruction has made great progress. \citet{13} used deep neural network for the first time to solve the problem of SR reconstruction and proposed SRCNN, which attracted extensive attention. Later, various SR reconstruction methods based on deep learning were proposed. At that time, the main optimization goal was to minimize the mean square error (MSE) of GT images and reconstructed images \citep{13,14} or maximize the peak signal-to-noise ratio (PSNR), which were also two metrics commonly used to evaluate and compare SR methods \citep{15}. However, the ability of MSE (or PSNR) to capture perceptual-related differences (e.g., texture details) is very limited. To solve this problem, \citet{16} proposed perceptual loss, which extracted image features through pre-trainied network to predict the uncertainty factors under LR. Subsequently, considering the great success of GAN \citep{18} in image generation, \citet{17} combined it with the perceptual-driven method, and made a further breakthrough in the clarity of the reconstructed image.
\par
Although perceptual loss has made great progress in improving visual quality, there are still some obvious defects. One of them is that only relying on a single pre-trained network cannot mine potential features of images throughly, which limits the reasoning ability of the network. In order to solve this problem, based on the ResNet network \citep{19}, this paper proposes a perceptual loss called ResNet loss different from the VGG loss \citep{16,17} in extraction way. Compared with the network structure of VGG \citep{20}, the ResNet network is composed of a large number of residual blocks \citep{19}, and it can retain more feature information and will not be lost with the increase of network layers. Some features extracted by the two distinct networks will exist in a complementary form, so this paper adopts the way of combining VGG loss with ResNet loss to improve the information acquisition ability of the overall perceptual features.
\par
In the process of optimizing VGG loss and ResNet loss simultaneously, we expect that them can both provide strong support for the network. However, the two losses are generated in different ways, so their magnitudes will be different. If they are added directly, the difference in magnitude will cause the network to be biased towards learning the perceptual features extracted by a single pre-trained network. As a result, the advantage of dual perceptual cannot be fully utilized. The use of static constants for weighting and combination to balance the magnitude also has some defects. It can only control the initial state of the two losses, but cannot ensure the two losses to continue to be at the same magnitude during the training process. To cope with the problem mentioned above, this paper proposes a dynamic weighting method. It takes the VGG loss as the reference target and weights the ResNet loss dynamically, which keeps the relative size of two losses stable. This method eliminates the influence of magnitude, and enables the network to learn both ResNet features and VGG features with a strong degree of attention. Therefore, we define the combination of the VGG loss and the ResNet loss with dynamically weighted as a dual perceptual loss (DP Loss). In addition, DP Loss has a certain degree of flexibility by using the dynamic weight, hence the DP Loss can be applied to different models conveniently.
\par
The hyperparameter setting of DP Loss has a great influence on the performance of the model. Therefore, we first apply DP Loss to SRGAN \citep{17} to get SRGAN with Dual Perceptual Loss (SRGAN-DP), and test the influence of different hyperparameter combinations on the model to obtain the optimal hyperparameter combination. According to the experimental results, the SRGAN-DP under the optimal condition is better than SRGAN in all the evaluation metrics. Then, we apply the DP Loss under the optimal hyperparameter combination to ESRGAN and get ESRGAN with Dual Perceptual Loss (ESRGAN-DP). Finally, the experimental results show that the ESRGAN-DP has better visual effect and evaluation metrics compared with SRGAN-DP.
\par
\begin{figure*}[htbp]
\centering
 \begin{minipage}{0.35\textwidth} % change this
 \centering
  \includegraphics[width=1\textwidth]{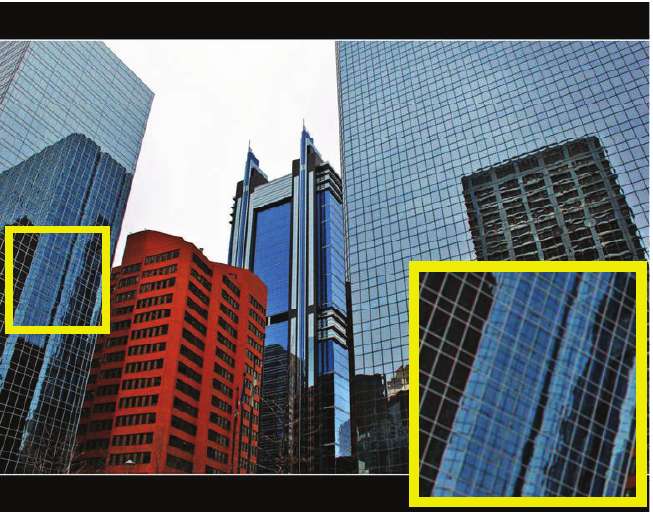}
  \subcaption*{a}
  \end{minipage}
\centering
 \begin{minipage}{0.45\textwidth} % change this
\centering
  \begin{minipage}{1\textwidth}
  \centering
  \begin{minipage}{0.3\textwidth}
  \centering
  \includegraphics[width=1\textwidth]{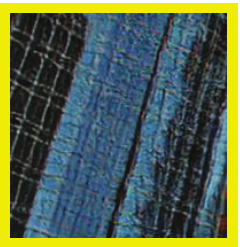}
  \subcaption*{b}
  \end{minipage}
  \begin{minipage}{0.3\textwidth}
  \centering
  \includegraphics[width=1\textwidth]{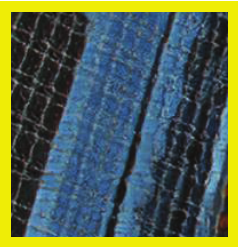}
  \subcaption*{c}
  \end{minipage}
  \begin{minipage}{0.3\textwidth}
  \centering
  \includegraphics[width=1\textwidth]{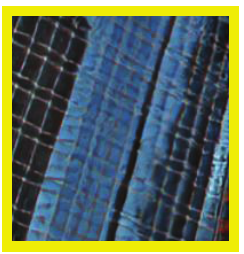}
  \subcaption*{d}
  \end{minipage}
  \end{minipage}

\centering
  \begin{minipage}{1\textwidth}
  \centering
  \begin{minipage}{0.3\textwidth}
  \centering
  \includegraphics[width=1\textwidth]{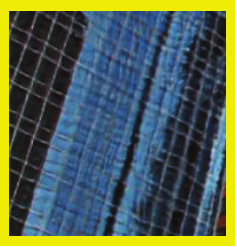}
  \subcaption*{e}
  \end{minipage}
  \begin{minipage}{0.3\textwidth}
  \centering
  \includegraphics[width=1\textwidth]{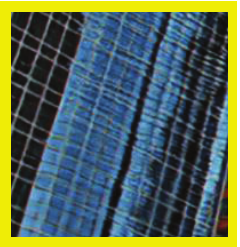}
  \subcaption*{f}
  \end{minipage}
  \begin{minipage}{0.3\textwidth}
  \centering
  \includegraphics[width=1\textwidth]{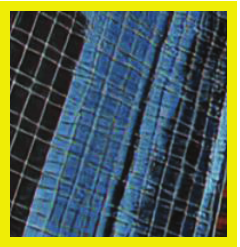}
  \subcaption*{g}
  \end{minipage}
  \end{minipage}
\end{minipage}

\captionsetup{labelfont={bf}}
\caption{Comparison between the proposed method and five state-of-the-art perceptual-driven SR methods based on GAN at 4$\times$ upscaling factor. The image ``Img099'' is from Urban100. (\textbf{a})Original. (\textbf{b})SRGAN. (\textbf{c})EnhanceNet. (\textbf{d})ESRGAN. (\textbf{e})SFTGAN. (\textbf{f})ESRGAN+. (\textbf{g})Ours}
\label{fig:1} 
\end{figure*}
\begin{figure*}[htbp]
\centering
  \includegraphics[width=0.9\textwidth]{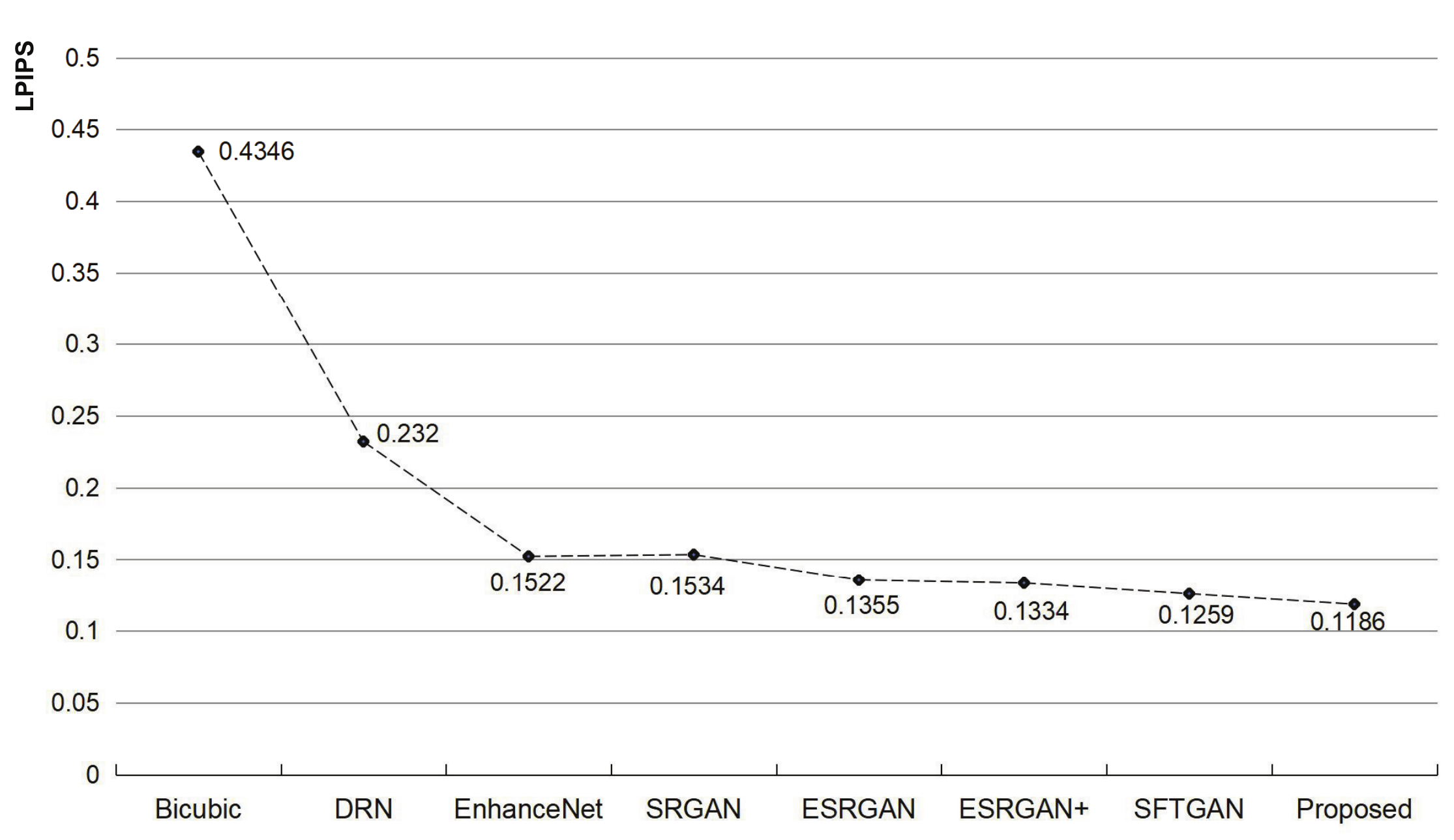}
\captionsetup{labelfont={bf}}
\caption{Comparison of LPIPS values of different SR methods on Urban100 dataset}
\label{fig:2}       % Give a unique label
\end{figure*}

The SR methods based on both perceptual-driven and GAN can acquire better visual effect. Fig. \ref{fig:1} shows the images reconstructed by ESRGAN-DP and those reconstructed by several SR methods mentioned above (SRGAN \citep{17}, EnhanceNet \citep{21}, ESRGAN \citep{22}, SFTGAN \citep{23}, ESRGAN+ \citep{50}) under the condition of a 4$\times$ upscaling factor. As shown in Fig. \ref{fig:1}, it is obviously that the lines on the surface of buildings are clearer and the reconstructed images have more realistic details after the DP Loss is applied to ESRGAN. The experimental results show that the performance of proposed method outperforms to other SR methods. Fig. \ref{fig:2} shows the comparison of DP Loss and the state-of-the-art SR methods in terms of LPIPS \citep{25} values on the Urban100. The smaller the value, the higher the perceptual similarity of the reconstructed image and the GT image. It can be seen from the line chart that the LPIPS value of ESRGAN-DP is the smallest among all SR methods. Therefore, it is concluded that the reconstruction effect of the image is effectively improved by adding DP Loss into the ESRGAN. 

\par
The main contributions of this paper are itemized as follows:
\\ 
\\
1. \quad Following the VGG loss, we propose the ResNet loss that is generated by the perceptual features extracted by the pre-trained ResNet network. Compared with VGG, the ResNet network does not lose the original information with the increase of layers, thus, a deeper level of output can be used to obtain the higher level of perceptual features.\\
2. \quad We propose the idea that the generator learns VGG features and ResNet features simultaneously. According to the differences of VGG and ResNet networks in forward calculation, it can be inferred that the features extracted by two networks are complementary, therefore, the recover ability of generator to texture details can be further improved.\\
3. \quad In this paper, the ResNet loss is dynamically weighted to eliminate the interference caused by the magnitude difference, so that both VGG loss and ResNet loss can bring greater benefits to the model.\\
4. \quad Considering the influence of the hyperparameter combination in DP Loss on the reconstruction results, we conduct experiments on different benchmark datasets. By comparing and analyzing the experimental results, the loss function under the optimal hyperparameter combination is selected and applied to ESRGAN. Experiments show that compared with the traditional ESRGAN, the visual effects and evaluation metrics have been significantly improved, and compared with other state-of-the-art methods are also outstanding.\\ 
\par
In Section \ref{sec:2}, we discuss the related network models and the research progress of loss functions. In Section \ref{sec:3}, we focus on the theoretical ideas related to DP Loss. In Section \ref{sec:4}, we integrate DP loss into SRGAN for hyperparameter analysis, and apply the loss function under the optimal hyperparameter combination to ESRGAN to obtain ESRGAN-DP, and then compare ESRGAN-DP with other state-of-the-art SR methods to verify the effectiveness of the proposed method; Section \ref{sec:5} concludes this paper.

\section{Related work}
\label{sec:2}
This section first introduces different network structures in the SR field, and then focuses on the improvement of related loss functions.
\par
Since the pioneering SRCNN \citep{13} was proposed, the use of deep learning to solve the SISR problem has attracted more and more attention. Compared with traditional methods, deep learning has obvious advantages. Not only does it have a great improvement in visual quality, but also has more diversity in optimization and improvement. Furthermore, more and more network architectures are used to solve the problem of SR reconstruction. \citet{26} proposed the VDSR, which increased the number of network layers to 20 and significantly improved the reconstruction effect. \citet{27} constructed the EDSR by using residual blocks that deleted unnecessary Batch Normalization (BN) layers \citep{28}. Inspired by DenseNet \citep{29}, \citet{30} proposed Residual Dense Block (RDB) and applied it to SR, and further discussed the deeper network structure with channel attention mechanism \citep{31}. \citet{22} proposed the Residual in Residual Dense Block (RRDB) without BN layer in the generative network. In order to restore more faithful textures, \citet{23} proposed SFTGAN, which used the prior category information for targeted generation. They constructed the Spatial Feature Transform (SFT) layer, and adjusted the features of the middle layer in a single network by using semantic segmentation maps \citep{32,33,34}. In order to fundamentally reduce the computational cost, \citet{35} applied the AdderNets \citep{36} to SR, and developed a learnable energy activation to adjust the feature distribution and refine the relevant details. Aiming at solving the problem of the uncertainty of image down-sampling methods in real scenes and the large solution space from LR to HR, \citet{37} proposed a dual regression scheme that can not only learn the mapping relationship from LR to HR, but also can estimate the down-sampling kernel.
\par
\citet{16} believed that only focusing on optimizing the MSE or PSNR of the pixel space ratio of the GT image and the reconstructed image would make the reconstructed image smooth. Therefore, they proposed the perceptual loss, which was to improve the reconstruction effect by minimizing the feature space error between the GT image and the reconstructed image. After that, \citet{17} applied the perceptual loss and the GAN \citep{18} to SR, and constituted the total loss of the generator by using the weighted combination of several loss components, including the adversarial loss, the content loss and the perceptual loss. The loss function proposed by \citet{21} consisted of four parts, namely the pixel-wise loss, the perceptual loss, the texture matching loss and the adversarial loss. \citet{22} proposed ESRGAN based on the idea of SRGAN, which used a variety of techniques to further improve the texture details of the reconstructed image. In terms of perceptual loss, they proposed to use the output before the activation of the convolutional layer to obtain more feature information, so that the object to be minimized was the error of the feature space before activation. As a result, a convincing effect was obtained on the texture details of the reconstructed image and they won the PIRM2018-SR Challenge \citep{40} champion. \citet{41} proposed a targeted perceptual loss on the basis of the labels of object, background and boundary, which made the network reconstruct the image from multiple perspectives and improved the overall effect of the image. Therefore, discussing the perceptual loss is crucial to the improvement of the reconstruction results, especially the facticity of texture details. In order to reduce the unnatural artifacts generated in the perceptual-driven method, this paper designs a novel perceptual loss function to achieve this goal.

\section{Methods}
\label{sec:3}
%Text with citations \cite{RefB} and \cite{RefJ}.
We solve the SR problem by training deep neural network. According to the theoretical idea proposed by \citet{13}, the optimization objective is as follows: 
\begin{equation}
\min _{\theta} \frac{1}{n} \sum_{i=1}^{n} L\left(G\left(I_{i}^{L R} ; \theta\right), I_{i}^{H R}\right),
\end{equation}
where $I_{i}^{L R} \in \mathbf{R}^{C \times H \times W}$ and  $I_{i}^{H R} \in \mathbf{R}^{C \times H \times W}$ represent the $i\mbox{-}th$ LR and HR sub-image pairs in the training set, respectively. $G\left(I_{i}^{L R} ; \theta\right)$ represents the up-sampling network. $\theta$ is the parameter to be optimized within the neural network. $L$ is the loss function.
\par
The definition of $L$ is crucial, and it is closely related to the reconstruction effect. According to the theoretical idea \citet{17} propose, we define several losses by comparing the differences between the reconstructed image and the GT image from different perspectives, and constitute the total loss through the weighted sum of every loss component. Here, $L$ can be represented as:
\begin{equation}
L=\lambda l_{{content }}+\eta l_{{adversarial }}+\gamma l_{D P}\label{total_loss},
\end{equation}
where $l_{{content }}$ is the content loss of the pixel-wise 1-norm distance between images reconstructed by the generator and GT images, $l_{ {adversarial }}$  is the loss caused by the application of Relativistic Average GAN (RaGAN) \citep{42}, $l_{ {DP }}$  is the DP Loss we propose. $\lambda$, $\eta$ and $\gamma$ are the coefficients of balancing different loss terms, respectively. In Section \ref{sec:3.2}, the DP Loss proposed in this paper will be introduced in detail.
\par
\subsection{Adversarial network structure}
\label{sec:3.1}
In order to enable a large number of features to be better learned, in the network structure, we select ESRGAN \citep{22} with strong learning ability. The model is composed of a generator network and a discriminator network. The two networks conduct the adversarial training through the alternate optimization way.
\par
The generator network is a residual network composed of a large number of residual blocks. In order to retain more feature information, the network does not contain the BN layer. We progressively enlarge the image by continuously adding up-sampling blocks with a 2$\times$ upscaling factor (if a 3$\times$ upscaling image is required, only a up-sampling block with a 3$\times$ upscaling factor is used). Each up-sampling block consists of three steps. Firstly, the input data is amplified by using the nearest neighbor interpolation \citep{6} with a 2$\times$ upscaling factor, and then it passes through a 3$\times$3 filter, and finally the output data of the filter is activated by LeakyReLU.
\par
The input image size of the discriminator network is fixed at 128$\times$128. The discrimination method uses the theoretical idea of RaD proposed in \citep{42}, and the output form is:
\begin{equation} 
D_{R a}\left(I^{H R}, I^{L R}\right)=\sigma\left(D\left(I^{H R}\right)-avg\left[D\left(G\left(I^{L R}\right)\right)\right]\right)\label{DRa},
\end{equation}
where $avg[\cdot]$ represents the operation of taking the average for the discriminant value of all fake data in the mini-batch. According to Eq. (\ref{DRa}), $D_{R a}$ is asymmetric, and the optimization direction will be substantially changed only by changing the parameter order and discriminant target value. Therefore, regardless of whether the optimization is the discrimination loss or the generation loss, the target network in this way can benefit from real data and fake data rather than the part of the data.

\subsection{Perceptual loss}
\label{sec:3.2}
\subsubsection{VGG loss and ResNet loss}
\label{sec:3.2.1}
\citet{17} proposed to define the VGG loss according to the ReLU activation layer of the pre-trained 19 layer VGG network. In order to obtain more feature information, \citet{22} redefined the VGG loss after the convolutional layer and before the activation layer. This paper uses the VGG loss defined in \citep{22}, that is, the L1 norm loss function is used to define the Manhattan Distance between the reconstructed image features and the GT image features:
\begin{equation}
l_{V G G / i, j}=\frac{1}{C_{i,j}W_{i, j} H_{i, j}} \sum_{z=1}^{C_{i, j}} \sum_{x=1}^{W_{i, j}} \sum_{y=1}^{H_{i, j}}\left|\Phi_{i, j}\left(G_{\theta_{G}}\left(I^{L R}\right)\right)_{x, y, z}-\Phi_{i, j}\left(I^{H R}\right)_{x, y, z}\right|\label{VGG_loss},
\end{equation}
where $\Phi_{i, j}$ represents features obtained by the $j\mbox{-}th$ convolution (before activation) before the $i\mbox{-}th$ maxpooling layer in the VGG network. $C_{i, j}$, $W_{i, j}$ and $H_{i, j}$ are the dimensions of their respective feature spaces in the VGG network.
\par
\begin{figure*}[htbp]
\centering
  \includegraphics[width=0.9\textwidth]{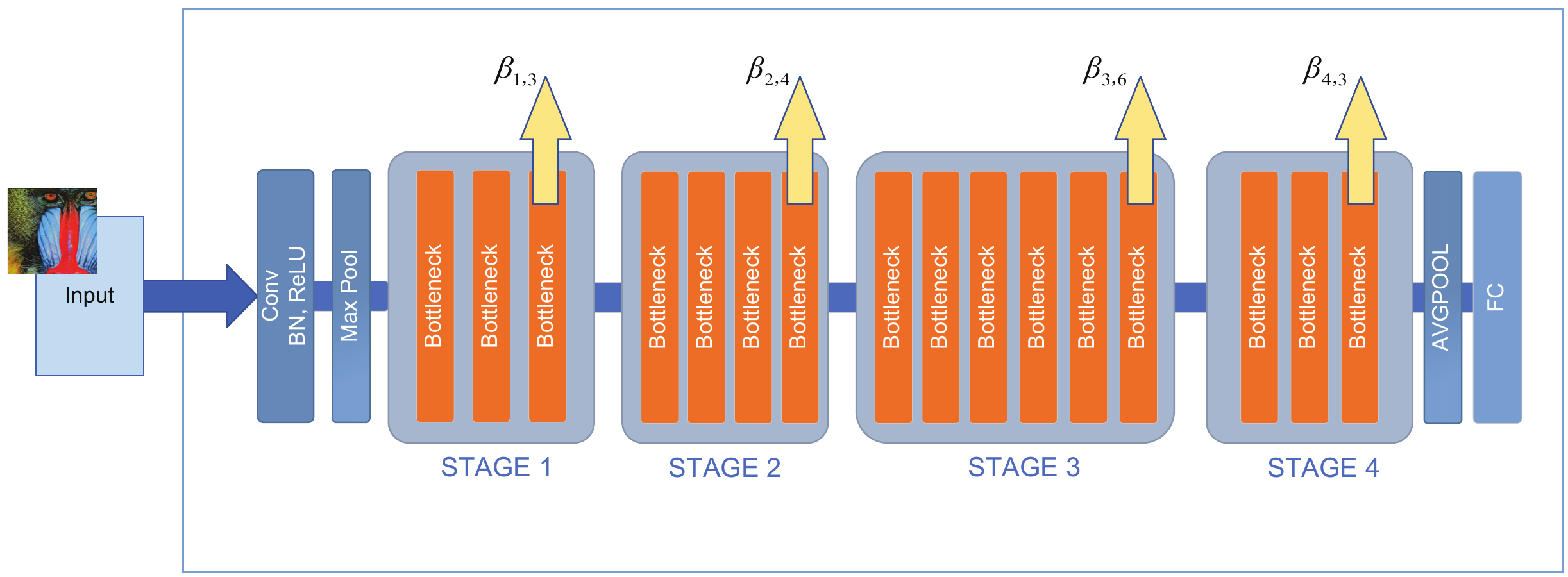}
\captionsetup{labelfont={bf}}
\caption{Block structure of the pre-trained 50 layer ResNet network}
\label{fig:3}       % Give a unique label
\end{figure*}

Based on the ideas of \citet{38}, \citet{39}, \citet{16} and \citet{17}, we define the ResNet loss on the ReLU activation layer of the pre-trained 50 layer ResNet network described in \citet{19}. Since the ResNet network is different from the VGG network in structure, we use a specific block way to specify each feature space. As shown in Fig. \ref{fig:3}, we divide ResNet-50 into four stages, each of which contains several bottleneck layers. The extracted perceptual features use the output value of the bottleneck layer at each stage, and the ResNet loss can also be expressed as:
\begin{equation}
l_{R E S / m, n}=\frac{1}{C_{m,n} W_{m, n} H_{m, n}} \sum_{z=1}^{C_{m, n}} \sum_{x=1}^{W_{m, n}} \sum_{y=1}^{H_{m, n}}\left|\beta_{m, n}\left(G_{\theta_{G}}\left(I^{L R}\right)\right)_{x, y, z}-\beta_{m, n}\left(I^{H R}\right)_{x, y, z}\right|\label{RES_loss},
\end{equation}
where $\beta_{m, n}$ represents features obtained by the $n\mbox{-}th$ bottleneck layer (after activation) at the $m\mbox{-}th$ stage. $C_{m, n}$, $W_{m, n}$ and $H_{m, n}$ are the dimensions of their respective feature spaces in the ResNet network. 

\par
In this paper, we adopt features of the last bottleneck layer at each stage to calculate ResNet loss, so there are four places to be tried in the 50 layer ResNet we use. But it is not necessarily that the features in deeper layers can bring better overall effect. The optimal feature map under specific conditions needs to be determined through experimental analysis of different situations.
\par
\begin{figure*}[htbp]
\centering
 \begin{minipage}{0.2\textwidth} % change this
 \centering
  \includegraphics[width=1\textwidth]{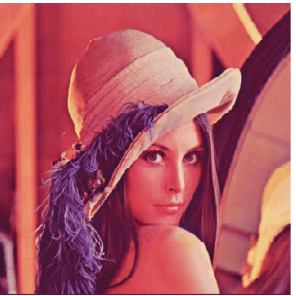}
  \subcaption*{a}
  \end{minipage}
\centering
 \begin{minipage}{0.55\textwidth} % change this
\centering
  \begin{minipage}{1\textwidth}
  \centering
  \begin{minipage}{0.3\textwidth}
  \centering
  \includegraphics[width=1\textwidth]{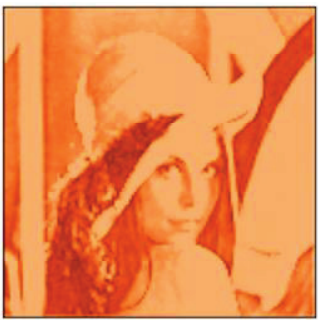}
  \subcaption*{b}
  \end{minipage}
  \begin{minipage}{0.3\textwidth}
  \centering
  \includegraphics[width=1\textwidth]{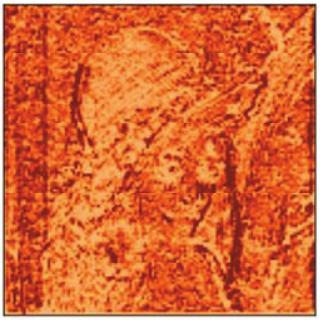}
  \subcaption*{c}
  \end{minipage}
  \begin{minipage}{0.3\textwidth}
  \centering
  \includegraphics[width=1\textwidth]{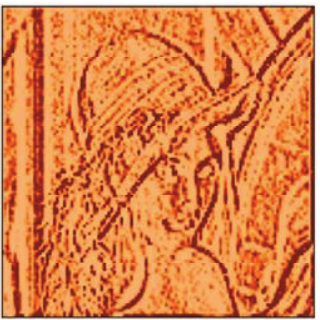}
  \subcaption*{d}
  \end{minipage}
  \end{minipage}

\centering
  \begin{minipage}{1\textwidth}
  \centering
  \begin{minipage}{0.3\textwidth}
  \centering
  \includegraphics[width=1\textwidth]{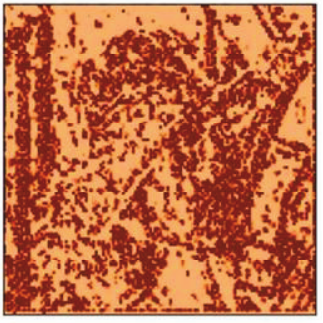}
  \subcaption*{e}
  \end{minipage}
  \begin{minipage}{0.3\textwidth}
  \centering
  \includegraphics[width=1\textwidth]{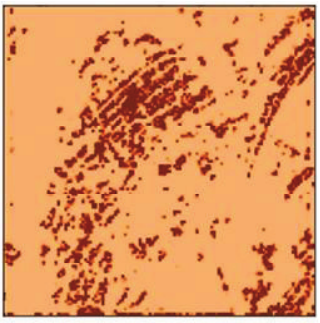}
  \subcaption*{f}
  \end{minipage}
  \begin{minipage}{0.3\textwidth}
  \centering
  \includegraphics[width=1\textwidth]{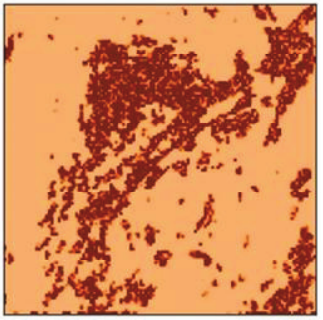}
  \subcaption*{g}
  \end{minipage}
  \end{minipage}
\end{minipage}

\captionsetup{labelfont={bf}}
\caption{Comparison of feature maps under different channels (after activation). The image ``Lenna'' is from Set14. (\textbf{a})Original. (\textbf{b})1st channel of $\beta_{1,2}$. (\textbf{c})64th channel of $\beta_{1,2}$. 
(\textbf{d})256th channel of $\beta_{1,2}$. (\textbf{e})1st channel of $\Phi_{3,3}$. (\textbf{f})64th channel of $\Phi_{3,3}$. (\textbf{g})256th channel of $\Phi_{3,3}$}
\label{fig:4} 
\end{figure*}

We can compare the differences and similarities of the perceptual features extracted by the pre-trained VGG network and the pre-trained ResNet network from a visual perspective. In order to make the features of two networks have certain comparability, we need to keep the length, width, and channel number of the two features as consistent as possible. The depth of the two networks that the images pass through is also as consistent as possible. In this paper, we choose $\beta_{1,2}$ and $\Phi_{3,3}$, which pass through two max-pooling layers and seven convolutions and are obtained after the ReLU activation layer. The length, width and the number of channels are exactly the same. The specific effect is shown in Fig. \ref{fig:4}. By visualizing the feature map under a single channel, it can be clearly seen that the perceptual ways of the two pre-trained networks are different. The features under $\Phi_{3,3}$ are more distinct, but some information is lost (e.g., the features under the 64th channel are very sparse). Benefiting from the excellent properties of the residual network, more original data is retained under $\beta_{1,2}$, especially the features under the 1st channel basically retains all single-channel pixel information, and the features under the 256th channel focuses on perceiving the edge of the image. This huge difference of perceptual ways is exactly what we need. 
\par
\begin{figure*}[htbp]
\centering
 \begin{minipage}{0.3\textwidth} % change this
 \centering
  \includegraphics[width=1\textwidth]{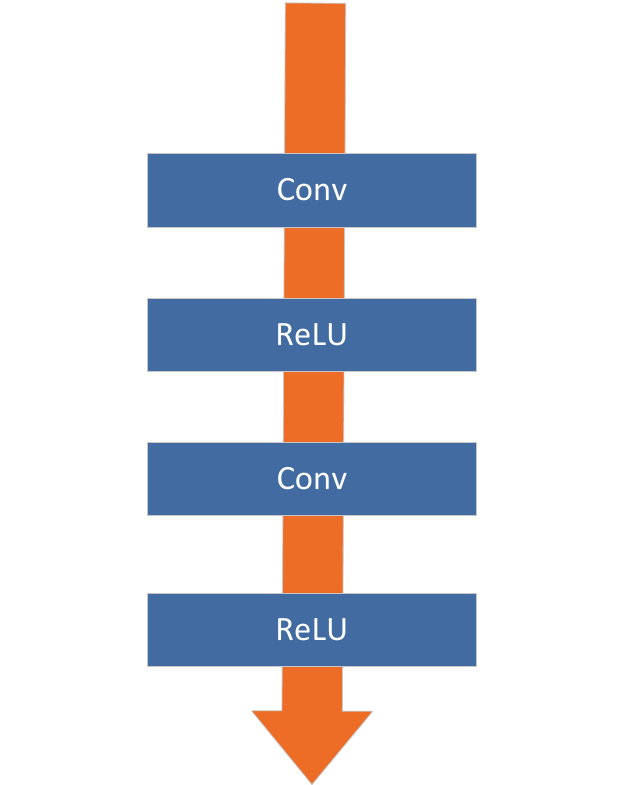}
  \subcaption*{a}
  \end{minipage}
 \begin{minipage}{0.3\textwidth} % change this
 \centering
  \includegraphics[width=1\textwidth]{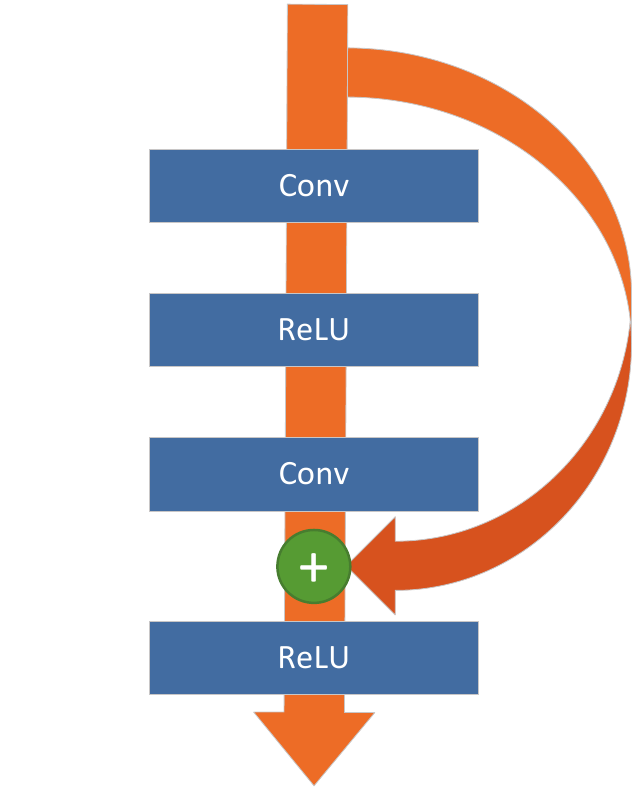}
  \subcaption*{b}
  \end{minipage}

\captionsetup{labelfont={bf}}

\caption{Comparison of basic connection way of the VGG network and ResNet network. (\textbf{a})VGG network. (\textbf{b})ResNet network}
\label{fig:5} 
\end{figure*}

In order to illustrate the source of the above-mentioned differences in a deeper level, we carry out the analysis from the forward calculation of the neural network, because the most essential difference between the two networks lies in the organization way of the network layer output. Fig. \ref{fig:5} shows the basic connection way of the VGG network and ResNet network. It can be seen that VGG uses multiple network layers to connect in sequence, while ResNet adopts a special jump connection way. The different connection ways make them have the distinct ways of forward calculation, which limits the back propagation, namely the optimization of network parameters. Furthermore, since the parameters of two networks are trained in the task of image recognition, some unimportant features will be discarded after the input features pass through the convolutional layer. Therefore, after the images pass through the VGG network, some features will become more obvious while other features will become weaker or even disappear with the number of layers increases. However, ResNet has the ability to retain the most original features while emphasizing important features due to its special forward calculation mechanism. Therefore, the features extracted by two networks have certain complementary properties. Fig. \ref{fig:4}(g) and Fig. \ref{fig:4}(d) can fully illustrate these two points, respectively. Fig. \ref{fig:4}(g) directly discards unimportant features (e.g., background) and highlights the most important feature information, while Fig. \ref{fig:4}(d) retains most of the original structural information and emphasizes edge position features.

\subsubsection{DP Loss}
\label{sec:3.2.2}
Obtaining more feature information from the image can enable the network to have better recovery capabilities in terms of faithful texture details. Therefore, we adopt a joint optimization strategy for the two perceptual losses mentioned in Section \ref{sec:3.2.1}. However, the differences in magnitude between the two losses bring some interference to the network training, which makes the advantages of dual perceptual not be exhibited. The details are as follows.
\par
Firstly, the loss function under the several perceptual losses is expressed as:
\begin{equation}
M P L=\sum_{i=1}^{n} P L_{i},
\end{equation}
where $P L$ represents a single perceptual loss function and $n$ represents the number of loss functions. Then, according to the gradient descent method, the network weight under the loss is updated, which can be expressed as:
\begin{equation}
w=w-\gamma \sum_{i=1}^{n} \frac{\partial P L_{i}}{\partial w}\label{muti_loss},
\end{equation}
where $w$ is the network weight, $\gamma$ is the learning rate. It can be seen from the Eq. (\ref{muti_loss}) that if there is a numerical difference in magnitude between $P L$s, there will also be such a difference in range of changes. Therefore, the gradient value of $w$ is limited by the magnitude, which leads to the fact that only a part of $P L$s occupy the dominant position. As a result, the advantages of dual perceptual cannot be exhibited. We can describe this problem through the extent to which the loss is sensitive to changes of network parameters. For example, we update the weight $w$: at a certain stage of the training process, $P L_{1}=1$, $P L_{2}=0.02$. Let $\Delta w \textgreater 0$, when $w + \Delta w$, then $P L_{1}=1.1$, $P L_{2}=0.01$; when $w - \Delta w$, then $P L_{1} = 0.9$, $P L_{2} = 0.03$. From above, it can be seen as follows: $\vert \Delta P L_{1} \vert=10 \ast \vert \Delta P L_{2} \vert$. If $ \Delta w \to 0 $, then $\frac{\partial P L_{1}}{\partial w}=-10 \ast \frac{\partial P L_{2}}{\partial w}$. The total gradient contains $\frac{\partial P L_{1}}{\partial w}$ and $\frac{\partial P L_{2}}{\partial w}$, so it is dominated by the gradient under $P L_{1}$ and its value is positive. However, in terms of the sensitivity of the loss, $P L_{2}$ is more sensitive to $\Delta w$ (comparing with itselves, $P L_{1}$ fluctuates by 0.1 times but $P L_{2}$ fluctuates by 0.5 times). Therefore, the total gradient is negative, that is, the gradient under $P L_{2}$ is dominant, which is more beneficial to the model. In summary, the difference of magnitude affects the correlation between the sensitivity of the loss and the total gradient, so that the model cannot have a strong attention to both perceptual features. Even in the later stages of the training process, $P L_{1}$ has converged, but the network still does not make too many adjustments for $P L_{2}$.
\par
In order to resolve the problems above, we eliminate the influence of the difference of magnitude by weighting $P L$s. In general, the static constant is used as the weight value, but the $P L$s in the training process is uncontrollable, which cannot ensure that the two losses are always at the same magnitude. In this paper, we use the dynamic weighting to solve this problem, and further define DP Loss $l_{D P}$. The specific formula is expressed as:
\begin{equation} 
l_{D P}=l_{V G G}+\frac{1}{\mu} \zeta_{l_{V G G}, l_{R E S}} l_{R E S}\label{DP_loss}, 
\end{equation}
where the ResNet loss $l_{R E S}$ is dynamically weighted and the weight value is $\frac{1}{\mu} \zeta_{l_{V G G}, l_{R E S}}$, $\mu$ is a nonzero constant. The $\zeta_{l_{V G G}, l_{R E S}}$ can be expressed as:
\begin{equation}
\zeta_{l_{V G G}, l_{R E S}}=value\left(\frac{l_{V G G}+c}{l_{R E S}+c}\right)\label{ratio},
\end{equation}
where $value\left(\frac{l_{V G G}+c}{l_{R E S}+c}\right)$ means to take the value of $\frac{l_{V G G}+c}{l_{R E S}+c}$ and disconnect the relationship between the functions associated with the value. $c$ is a tiny positive constant, so it can be negligible, and its role is just to keep the denominator from being zero. Therefore, $\frac{1}{\mu} \zeta_{l_{V G G}, l_{R E S}}$  is just a value that changes with the change of the ratio of $l_{V G G}$ to $l_{R E S}$. Thus, $\frac{1}{\mu} \zeta_{l_{V G G}, l_{R E S}}$ is taken as the weight value under the ResNet loss, which can only change the update range of network parameters, not the update direction.
\par
From Eq. (\ref{DP_loss}) and Eq. (\ref{ratio}), it can be seen that the weighted ResNet loss value always exists in DP Loss in the form of multiple of VGG loss value, and $\mu$ is used to determine the multiple. In the process of updating network, when the VGG loss becomes smaller, the ResNet loss will be forced to decrease, and vice versa. The purpose of this is to ensure that the relative size of the loss value between the VGG loss and the ResNet loss is in a fixed state, and the network is always trained in this state. Therefore, it is prevented the situation that the network focuses on learning the perceptual features extracted by a single network due to the difference of magnitude  between the perceptual losses during the training process, so that the advantages of dual perceptual can be fully utilized.
\par
The method proposed in this paper only performs dynamic weights for the perceptual loss. The purpose of this is to ensure that DP Loss has a certain degree of flexibility and is convenient for migration to other models. The specific process of obtaining DP Loss is described by algorithm \ref{alg:dp}. In addition, it is necessary to discuss the effect of the different combinations of $\mu$, $\Phi$ and $\beta$ in DP Loss on the model, this occurs because the effects under various conditions are very different in both the visual quality of the image and the evaluation metric. In Section \ref{sec:4.3}, we compare and analyze the hyperparameter combinations under different conditions to determine the optimal hyperparameter of DP Loss in the relevant model.

\begin{algorithm}

  \caption{Process of obtaining DP Loss}\label{alg:dp}
  \SetKwFunction{Value}{Value}
  \SetKwFunction{TruncationVGG}{TruncationVGG}\SetKwFunction{TruncationResNet}{TruncationResNet}
  \KwIn{The reconstructed image $I^{SR}$, the high-resolution image $I^{HR}$, the $\mu$, the $i$ and $j$ of $\Phi_{i,j}$, and the $m$ and $n$ of $\beta_{m,n}$.}
  \KwOut{DP Loss $l_{D P}$}
  \BlankLine
  \textbf{Initialization}:\\
  Load the pre-trained parameters for VGG19 and ResNet50 networks\;
  Intercept the VGG19 network to obtain $\TruncationVGG(\cdot)$ according to $i$ and $j$\;
  Intercept the ResNet50 network to obtain $\TruncationResNet(\cdot)$ according to $m$ and $n$\;
  
  \textbf{1)Obtain VGG loss}:\\
  Extract $I_{SR}$'s VGG features $P_{VGG}^{SR} \leftarrow \TruncationVGG(I_{SR})$\;
  Extract $I_{HR}$'s VGG features $P_{VGG}^{HR} \leftarrow \TruncationVGG(I_{HR})$\;
  Calculate the Manhattan Distance between  $P_{VGG}^{SR}$ and $P_{VGG}^{HR}$ to obtain VGG loss $l_{VGG}$\;
  \textbf{2)Obtain ResNet loss}:\\
  Extract $I_{SR}$'s ResNet features $P_{RES}^{SR} \leftarrow \TruncationResNet(I_{SR})$\;
  Extract $I_{HR}$'s ResNet features $P_{RES}^{HR} \leftarrow \TruncationResNet(I_{HR})$\;
  Calculate the Manhattan Distance between  $P_{RES}^{SR}$ and $P_{RES}^{HR}$ to obtain ResNet loss $l_{RES}$\;
  \textbf{3)Obtain DP Loss}:\\
  Calculate the ratio $R_{l_{VGG},l_{RES}}$ of VGG loss to ResNet loss, where $R_{l_{VGG},l_{RES}} = \frac{l_{VGG}}{l_{RES}}$\;
  Break off the relationship between the functions associated with the $R_{l_{VGG},l_{RES}}$ to obtain $\zeta_{l_{V G G}, l_{R E S}}$\;
  Calculate DP Loss $ l_{D P} \leftarrow l_{V G G}+\frac{1}{\mu} \zeta_{l_{V G G}, l_{R E S}} l_{R E S}$\; 

\end{algorithm}

\section{Experiments}
\label{sec:4}
\subsection{Datasets and evaluation metrics}
\label{sec:4.1}
The training set includes 800 high-definition images from the public dataset DIV2K \citep{44}. By using sliding window to crop the images, 32,592 non-overlapping sub-images with the size of 480$\times$480 are obtained. And then we perform bicubic interpolation operation on these images to get the corresponding downsampling images. Following ESRGAN, only a 4$\times$ upscaling factor is considered in the experiment. We take the widely used benchmark datasets as our test datasets. The dataset contains Set5 \citep{45}, Set14 \citep{11}, BSD100 \citep{24} and Urban100 \citep{46}, which have 5 images, 14 images, 100 images and 100 images, respectively.
\par
The peak signal-to-noise ratio (PSNR), structural similarity (SSIM) \citep{47} and learned perceptual image patch similarity (LPIPS) \citep{25} are evaluation metrics in our experiment. The lower the LPIPS value, the higher the perceptual similarity, that is, the reconstructed images are closer to GT images in visual quality. In order to make a fair comparison, a 4-pixel wide stripe is removed from each border of all the evaluated images, and the Y-channel of the images is used to calculate PSNR and SSIM.
\subsection{Training details}
\label{sec:4.2}
We use a GTX 2080Ti GPU and apply the pytorch \citep{48} framework to train and test all models. The setting of experimental parameters follows \citep{22}. Herein, the mini-batch size is set to 16, the spatial size of the cropped HR patch is 128$\times$128 and parameters of the generator loss function in Eq. (\ref{total_loss}) are $\lambda=1e-2$, $\eta=5e-3$ and $\gamma=1$. The model training needs 400K iterations in total. The initial learning rate is $1e-2$, and the learning rate is halved when the iteration reaches [50k, 100k, 200k, 300k]. The Adam \citep{49} with $\beta_{1}=0.9$ and $\beta_{2}=0.99$ is used to alternately optimize the generator and the discriminator.

\subsection{Hyperparameter analysis}
\label{sec:4.3}
According to Eq. (\ref{VGG_loss}), (\ref{RES_loss}) and (\ref{DP_loss}), it can be seen that there are three important hyperparameters need to be specified, which are the VGG features $\Phi$, the ResNet features $\beta$ and the constant $\mu$. For the $\Phi$, we refer to \citep{17,22} and designate it as $\Phi_{5,4}$ (before activation), while $\beta$ and $\mu$ need to be determined according to the experiment. In this paper, we set the optional hyperparameters of $\beta$ as $\beta_{1,3}$, $\beta_{2,4}$, $\beta_{3,6}$ and $\beta_{4,3}$ (after activation), and the optional hyperparameters of $\mu$ as 0.2, 0.5, 1, 5, 10 and 20. Since SRGAN and ESRGAN are similar in implementation ideas, it can be proved that there is a certain positive correlation in the improvement of the image reconstruction effect of the two models using different hyperparameter combinations (the correlation will be proved in Section \ref{sec:4.3.4}). Therefore, in order to reduce the training burden, we use SRGAN to conduct comparative experiments under each hyperparameter combination, and apply the obtained optimal hyperparameter combination to ESRGAN. In order to improve the efficiency of finding the optimal hyperparameters, we obtain the optimal hyperparameter combination by alternately fixing a certain hyperparameter. In order to ensure that the results are fair and effective, the number of iterations is 400K during training.

\subsubsection{Determination of hyperparameter $\mu$}
\label{sec:4.3.1}
\begin{figure*}[htbp]
\centering
 \begin{minipage}{0.3\textwidth} % change this
 \centering
  \includegraphics[width=1\textwidth]{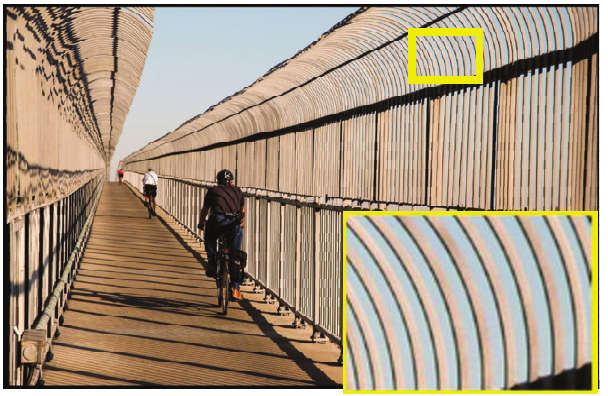}
  \subcaption*{a}
  \end{minipage}

 \begin{minipage}{1\textwidth} % change this
  \begin{minipage}{0.13\textwidth}
  \centering
  \includegraphics[width=1\textwidth]{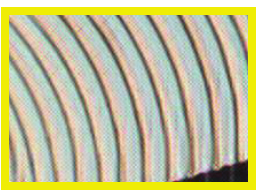}
  \subcaption*{b}
  \end{minipage}
 \centering
  \begin{minipage}{0.13\textwidth}
  \centering
  \includegraphics[width=1\textwidth]{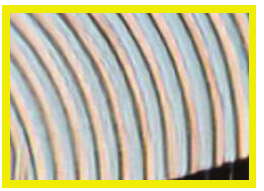}
  \subcaption*{c}
  \end{minipage}
  \begin{minipage}{0.13\textwidth}
  \centering
  \includegraphics[width=1\textwidth]{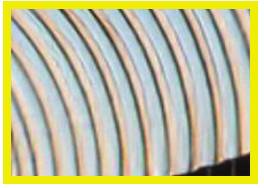}
  \subcaption*{d}
  \end{minipage}
  \begin{minipage}{0.13\textwidth}
  \centering
  \includegraphics[width=1\textwidth]{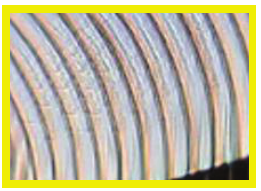}
  \subcaption*{e}
  \end{minipage}
  \begin{minipage}{0.13\textwidth}
  \centering
  \includegraphics[width=1\textwidth]{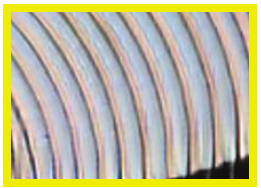}
  \subcaption*{f}
  \end{minipage}
  \begin{minipage}{0.13\textwidth}
  \centering
  \includegraphics[width=1\textwidth]{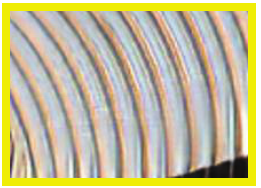}
  \subcaption*{g}
  \end{minipage}
  \begin{minipage}{0.13\textwidth}
  \centering
  \includegraphics[width=1\textwidth]{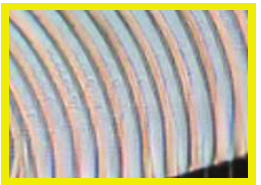}
  \subcaption*{h}
  \end{minipage}
\end{minipage}
  \centering
  \begin{minipage}{0.9\textwidth}
  \centering
  \includegraphics[width=1\textwidth]{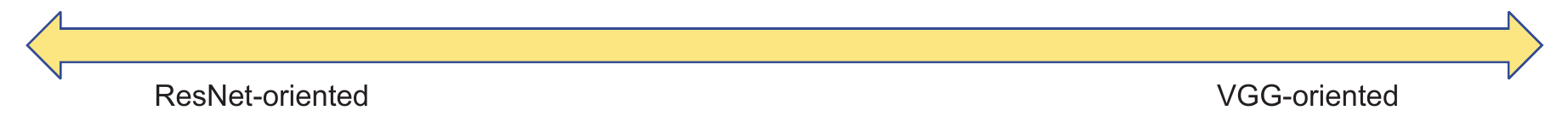}
  \end{minipage}

\captionsetup{labelfont={bf}}
\caption{Comparison of visual effects of DP Loss under different $\mu$ . The image ``Img024'' is from Urban100. (\textbf{a})Original. (\textbf{b})$\mu=0.2$. (\textbf{c})$\mu=0.5$. (\textbf{d})$\mu=1$. (\textbf{e})$\mu=5$. (\textbf{f})$\mu=10$. (\textbf{g})$\mu=20$. (\textbf{h})$\mu=\infty$}
\label{fig:6} 
\end{figure*}

We fix $\beta$ and set it to $\beta_{1,3}$ firstly, and then compare the experimental results under the different values of $\mu$. The experimental results on the visual effect is shown in Fig. \ref{fig:6}. When the $\mu$ is 0.5 or 1, the lines become clearer and the textures are closer to the GT image. Table \ref{tab:1} gives the comparison of results caused by losses under the different values of $\mu$, and all evaluation metrics have the better performance after adding DP Loss under the appropriate value of $\mu$. When $\mu$ is 1, the values of SSIM and PSNR in each dataset are the best, followed by the values of LPIPS; when $\mu$ is 0.5, the values of LPIPS are the best, followed by most of the values of SSIM and PSNR. In this paper, we pay more attention to the LPIPS, namely the perceptual similarity. Compared with SSIM and PSNR, LPIPS is more in line with human perceptual habits. Finally, the value of $\mu$ is determined as 0.5.

\subsubsection{Determination of hyperparameter $\beta$}
\label{sec:4.3.2}
\begin{figure*}[htbp]
\centering
 \begin{minipage}{0.2\textwidth} % change this
 \centering
  \includegraphics[width=1\textwidth]{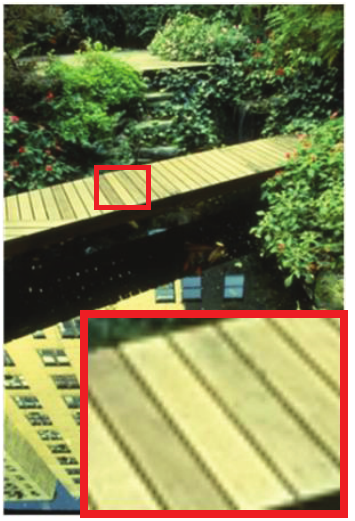}
  \subcaption*{a}
  \end{minipage}
 \begin{minipage}{0.65\textwidth} % change this
  \begin{minipage}{1\textwidth}
  \centering
  \begin{minipage}{0.23\textwidth}
  \centering
  \includegraphics[width=1\textwidth]{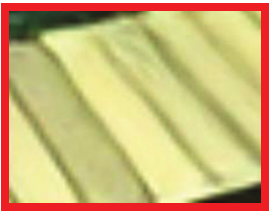}
  \subcaption*{b}
  \end{minipage}
  \begin{minipage}{0.23\textwidth}
  \centering
  \includegraphics[width=1\textwidth]{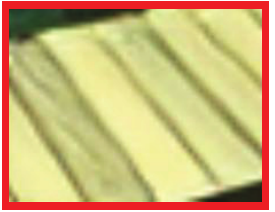}
  \subcaption*{c}
  \end{minipage}
  \begin{minipage}{0.23\textwidth}
  \centering
  \includegraphics[width=1\textwidth]{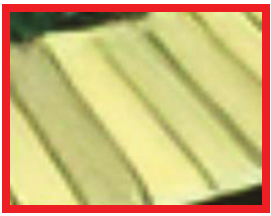}
  \subcaption*{d}
  \end{minipage}
  \begin{minipage}{0.23\textwidth}
  \centering
  \includegraphics[width=1\textwidth]{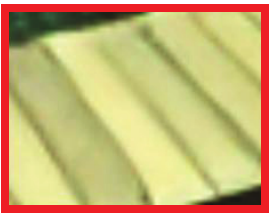}
  \subcaption*{e}
  \end{minipage}
  \end{minipage}
\end{minipage}

\captionsetup{labelfont={bf}}
\caption{Comparison of visual effects of DP Loss under different $\beta$. The image ``148026'' is from BSD100. (\textbf{a})Original. (\textbf{b})$\beta_{1,3}$. (\textbf{c})$\beta_{2,4}$. (\textbf{d})$\beta_{3,6}$. (\textbf{e})$\beta_{4,3}$ }
\label{fig:7} 
\end{figure*}

We use the optimal $\mu$ obtained in Section \ref{sec:4.3.1} as a fixed value, and then observe the influence on the image reconstruction when $\beta$ is under the different values. Comparisons of visual effects and objective evaluation metrics are shown in Fig. \ref{fig:7} and Table \ref{tab:2}, respectively. It can be seen that the result of DP Loss under the condition of $\beta_{3,6}$ occupies more advantages. From visual effects, the images reconstructed by using $\beta_{3,6}$ have less unnatural artifacts, so that the reconstructed images are closer to the GT images in structure. From evaluation metrics, it can be found that both PSNR and SSIM occupy the first or second best, and the LPIPS is the best in BSDS100 and Urban100 datasets. To sum up, the $\beta_{3,6}$ will be the optimal features.

\begin{table}[htbp]
\captionsetup{labelfont={bf}}
\caption{Comparison of results of DP Loss under the different $\mu$}
\label{tab:1}       % Give a unique label
\begin{threeparttable}
\resizebox{\textwidth}{!}{

	\begin{tabular}{lcccc}
	\toprule[1pt]
	\multirow{2}{*}{Hyperparameter} & \multicolumn{1}{l}{Set5} & \multicolumn{1}{l}{Set14} & \multicolumn{1}{l}{BSD100} & \multicolumn{1}{l}{Urban100}\\
	\cmidrule(lr){2-2} 	\cmidrule(lr){3-3} \cmidrule(lr){4-4} \cmidrule(lr){5-5} %画线
	&SSIM / PSNR / LPIPS & SSIM / PSNR / LPIPS & SSIM / PSNR / LPIPS & SSIM / PSNR / LPIPS \\
	\midrule
	$\mu=\infty$ & 0.7706 / 26.89 / 0.1085 & 0.7347 / 26.07 / 0.1239 & 0.6919 / 25.37 / 0.1483 & 0.7150 / 24.61 / 0.1407  \\
	$\mu=0.2$ & 0.3596 / 20.33 / 0.1500 & 0.3470 / 20.12 / 0.1672 & 0.3075 / 19.81 / 0.2089 & 0.3443 / 19.64 / 0.1830 \\
	$\mu=0.5$ & {\color{blue}0.7922} / {\color{blue}27.39} / {\color{red}0.1052} & {\color{blue}0.7570} / 26.37 / {\color{red}0.1207} & {\color{blue}0.7225} / 25.82 / {\color{red}0.1392} & {\color{blue}0.7460} / {\color{blue}25.07} / {\color{red}0.1297} \\
	$\mu=1$ & {\color{red}0.7959} / {\color{red}27.47} / {\color{blue}0.1080} & {\color{red}0.7617} / {\color{red}26.51} / {\color{blue}0.1226} & {\color{red}0.7259} / {\color{red}25.95} / {\color{blue}0.1447} & {\color{red}0.7490} / {\color{red}25.18} / {\color{blue}0.1331} \\
	$\mu=5$ & 0.7905 / 27.37 / 0.1092 & 0.7565 / {\color{blue}26.49} / 0.1236 & 0.7188 / {\color{blue}25.86} / 0.1489 & 0.7391 / 25.04 / 0.1386 \\
	$\mu=10$ & 0.7826 / 27.13 / 0.1082 & 0.7486 / 26.28 / 0.1230 & 0.7111 / 25.71 / 0.1473 & 0.7311 / 24.87 / 0.1393 \\
    $\mu=20$ & 0.7814 / 27.13 / 0.1098 & 0.7481 / 26.35 / 0.1250 & 0.7090 / 25.68 / 0.1495 & 0.7288 / 24.86 / 0.1413 \\
	\bottomrule[1pt]
	\end{tabular}

}
      \begin{tablenotes} %添加此处
		\item $\mu=\infty$ means only using the VGG loss, namely the original perceptual loss.
		\item The best performance is highlighted in {\color{red}red } (1st best) and {\color{blue} blue} (2nd best).
      \end{tablenotes} %添加此处
\end{threeparttable} %添加此处
\end{table}

\begin{table}[H]
\captionsetup{labelfont={bf}}
\caption{Comparison of DP Loss under different $\beta$}
\label{tab:2}       % Give a unique label
\begin{threeparttable}
\resizebox{\textwidth}{!}{
	\begin{tabular}{ccccc}
	\toprule[1pt]
	\multirow{2}{*}{Hyperparameter} & \multicolumn{1}{l}{Set5} & \multicolumn{1}{l}{Set14} & \multicolumn{1}{l}{BSD100} & \multicolumn{1}{l}{Urban100}\\
	\cmidrule(lr){2-2} 	\cmidrule(lr){3-3} \cmidrule(lr){4-4} \cmidrule(lr){5-5} %画线
	&SSIM / PSNR / LPIPS & SSIM / PSNR / LPIPS & SSIM / PSNR / LPIPS & SSIM / PSNR / LPIPS \\
	\midrule
	$\beta_{1,3}$ & {\color{red}0.7922} / {\color{blue}27.39} / 0.1052 & {\color{red}0.7570} / {\color{blue}26.37} / 0.1207 & {\color{red}0.7225} / {\color{red}25.82} / 0.1392 & {\color{red}0.7460} / {\color{red}25.07} / 0.1297  \\
	$\beta_{2,4}$ & 0.7899 / 27.27 / {\color{red}0.1039} & 0.7547 / 26.26 / {\color{blue}0.1187} & {\color{blue}0.7152} / 25.59 / {\color{blue}0.1371} & 0.7406 / 24.90 / {\color{blue}0.1283} \\
	$\beta_{3,6}$ & {\color{blue}0.7915} / {\color{red}27.45} / 0.1051 & {\color{blue}0.7564} / {\color{red}26.46} /  0.1200 & {\color{blue}0.7202} / {\color{red}25.82} / {\color{red}0.1370} & {\color{blue}0.7434} / {\color{blue}25.06} / {\color{red}0.1278} \\
	$\beta_{4,3}$ & 0.7876 / 27.13 / {\color{blue}0.1040} & 0.7543 / 26.30 / {\color{red}0.1176} & 0.7131 / 25.60 / 0.1405 & 0.7372 / 24.89 / 0.1327 \\
	\bottomrule[1pt]
	\end{tabular}
}
      \begin{tablenotes} %添加此处
		\item The best performance is highlighted in {\color{red}red } (1st best) and {\color{blue} blue} (2nd best).
      \end{tablenotes} %添加此处
\end{threeparttable} %添加此处
\end{table}

\subsubsection{SRGAN and ESRGAN under the DP Loss with the optimal hyperparameter}
\label{sec:4.3.3}
We use the two optimal hyperparameters in DP Loss obtained from Section \ref{sec:4.3.1} and Section \ref{sec:4.3.2} and apply them to SRGAN and ESRGAN respectively to obtain SRGAN-DP and ESRGAN-DP. The specific effects are given in Table \ref{tab:3}. Compared with SRGAN-DP, ESRGAN-DP has a more superior effect in terms of LPIPS.

\begin{table}[H]
\captionsetup{labelfont={bf}}
\caption{Comparison between SRGAN-DP and ESRGAN-DP in LPIPS values}
\label{tab:3}       % Give a unique label
\begin{threeparttable}
	\begin{tabular}{lcccc}
	\toprule[1pt]
	\multirow{2}{*}{Methods} & \multicolumn{4}{c}{DataSet}\\
	\cmidrule(lr){2-5} %画线
	&Set5 & Set14& BSD100 & Urban100 \\
	\midrule
	SRGAN-DP & 0.1051 & 0.1200 & 0.1370 & 0.1278 \\
	ESRGAN-DP & \textbf{0.0990} & \textbf{0.1139} & \textbf{0.1280} & \textbf{0.1186} \\
	\bottomrule[1pt]
	\end{tabular}

      \begin{tablenotes} %添加此处
		\item \textbf{Font bold} indicates the best performance.
      \end{tablenotes} %添加此处
\end{threeparttable} %添加此处
\end{table}

\subsubsection{The positive correlation between SRGAN-DP and ESRGAN-DP}
\label{sec:4.3.4}
In order to prove whether there is a certain positive correlation between the effects of SRGAN-DP and ESRGAN-DP under different hyperparameter combinations, we select four combinations of representative hyperparameters from Sections \ref{sec:4.3.1} and \ref{sec:4.3.2}, which are $\mu=\infty$, $\mu=1$ + $\beta_{1,3}$, $\mu=10$ + $\beta_{1,3}$ and $\mu=0.5$ + $\beta_{3,6}$, respectively. In SRGAN-DP, the performance effects of the four hyperparameter combinations are: when $\mu=\infty$, all evaluation metrics are the worst; when $\mu=1$ + $\beta_{1,3}$, PSNR and SSIM are the best; when $\mu=10$ + $\beta_{1,3}$, each metric is neither the best nor the worst; when $\mu=0.5$ + $\beta_{3,6}$, LPIPS is the best. It can be seen from Table \ref{tab:4} that the effects shown in SRGAN-DP are also shown in ESRGAN-DP. Therefore, it can be concluded that the effects of SRGAN-DP and ESRGAN-DP under different hyperparameter combinations have a certain positive correlation.

\begin{table}[htbp]
\captionsetup{labelfont={bf}}
\caption{Comparison of different hyperparameter combinations between SRGAN-DP and ESRGAN-DP on BSD100}
\label{tab:4}       % Give a unique label
\begin{threeparttable}
	\begin{tabular}{lcccccc}
	\toprule[1pt]
	\multirow{2}{*}{Hyperparameter } & \multicolumn{3}{c}{SRGAN-DP} & \multicolumn{3}{c}{ESRGAN-DP}\\
	\cmidrule(lr){2-4} 	\cmidrule(lr){5-7} %画线
	&PSNR & SSIM & LPIPS & PSNR & SSIM & LPIPS \\
	\midrule
	$\mu=\infty$ & \underline{25.37} & \underline{0.6919} & \underline{0.1483} & \underline{24.95} & \underline{0.6785} & \underline{0.1428}  \\
	$\mu=1$ + $\beta_{1,3}$ & \textbf{25.95} & \textbf{0.7259} & 0.1447 & \textbf{25.43} & \textbf{0.7007} & 0.1329\\
	$\mu=10$ + $\beta_{1,3}$ & 25.71 & 0.7111 & 0.1473 & 25.35 & 0.6968 & 0.1403 \\
	$\mu=0.5$ + $\beta_{3,6}$ & 25.82 & 0.7202 & \textbf{0.1370} & 25.40 & 0.6993 & \textbf{0.1280} \\
	\bottomrule[1pt]
	\end{tabular}
      \begin{tablenotes} %添加此处
		\item The best performance is highlighted in \textbf{bold} and the worst performance is highlighted in \underline{underline}.
      \end{tablenotes} %添加此处
\end{threeparttable} %添加此处
\end{table}

\subsection{Comparison with the state-of-the-art technologies}
\label{sec:4.4}
To demonstrate the effectiveness of ESRGAN-DP, we compare it with the state-of-the-art single image SR methods in terms of evaluation metrics and visual quality, including EnhanceNet \citep{21}, SRGAN \citep{17}, ESRGAN \citep{22}, SFTGAN \citep{23} DRN \citep{37} and ESRGAN+ \citep{50}.
\par
\begin{figure*}[htbp]
\centering
 \begin{minipage}{1\textwidth} % change this
  \begin{minipage}{1\textwidth}
  \centering
  \begin{minipage}{0.3\textwidth}
  \centering
  \includegraphics[width=1\textwidth]{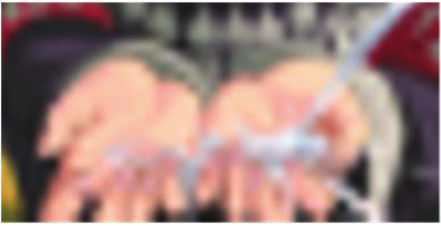}
  \subcaption*{a}
  \end{minipage}
  \begin{minipage}{0.305\textwidth}
  \centering
  \includegraphics[width=1\textwidth]{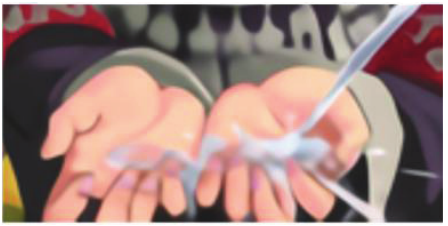}
  \subcaption*{b}
  \end{minipage}
  \begin{minipage}{0.3\textwidth}
  \centering
  \includegraphics[width=1\textwidth]{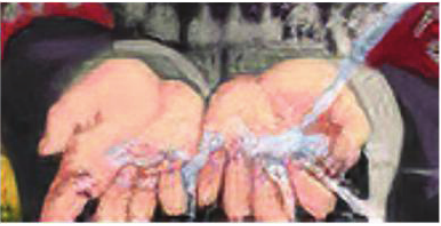}
  \subcaption*{c}
  \end{minipage}
  \end{minipage}

  \begin{minipage}{1\textwidth}
  \centering
  \begin{minipage}{0.3\textwidth}
  \centering
  \includegraphics[width=1\textwidth]{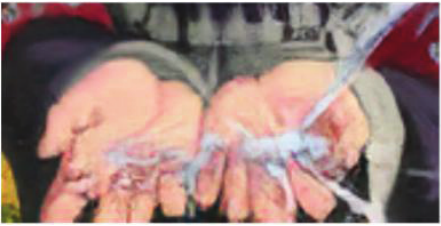}
  \subcaption*{d}
  \end{minipage}
  \begin{minipage}{0.3\textwidth}
  \centering
  \includegraphics[width=1\textwidth]{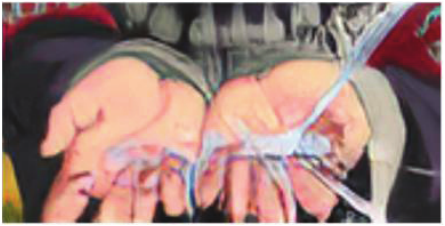}
  \subcaption*{e}
  \end{minipage}
  \begin{minipage}{0.3\textwidth}
  \centering
  \includegraphics[width=1\textwidth]{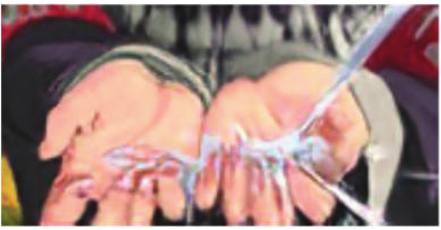}
  \subcaption*{f}
  \end{minipage}
  \end{minipage}

  \begin{minipage}{1\textwidth}
  \centering
  \begin{minipage}{0.3\textwidth}
  \centering
  \includegraphics[width=1\textwidth]{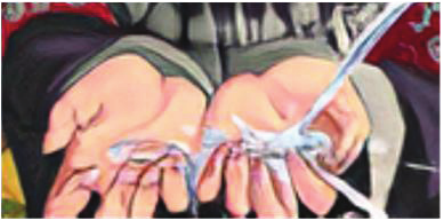}
  \subcaption*{g}
  \end{minipage}
  \begin{minipage}{0.3\textwidth}
  \centering
  \includegraphics[width=1\textwidth]{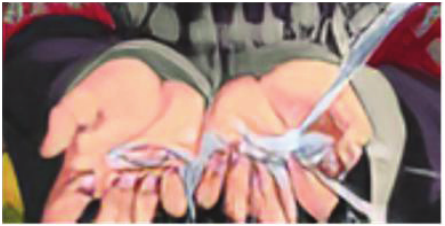}
  \subcaption*{h}
  \end{minipage}
  \begin{minipage}{0.3\textwidth}
  \centering
  \includegraphics[width=1\textwidth]{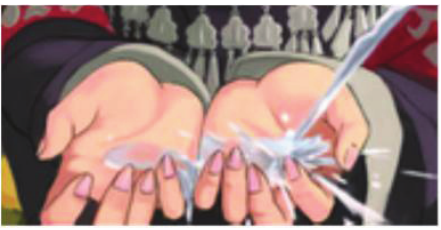}
  \subcaption*{i}
  \end{minipage}
  \end{minipage}
\end{minipage}

\captionsetup{labelfont={bf}}
\caption{Comparison of results of different SR methods on ``Comic" image from Set14. (\textbf{a})Bicubic. (\textbf{b})DRN. (\textbf{c})EnhanceNet. (\textbf{d})SRGAN. (\textbf{e})ESRGAN. (\textbf{f})SFTGAN. (\textbf{g})ESRGAN+. (\textbf{h})Ours. (\textbf{i})Original}
\label{fig:8} 
\end{figure*}
\begin{figure*}[htbp]
\centering
 \begin{minipage}{1\textwidth} % change this
  \begin{minipage}{1\textwidth}
  \centering
  \begin{minipage}{0.3\textwidth}
  \centering
  \includegraphics[width=1\textwidth]{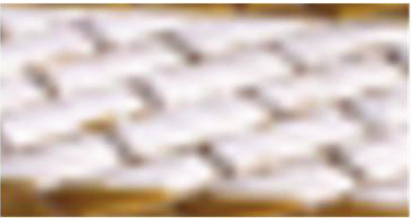}
  \subcaption*{a}
  \end{minipage}
  \begin{minipage}{0.3\textwidth}
  \centering
  \includegraphics[width=1\textwidth]{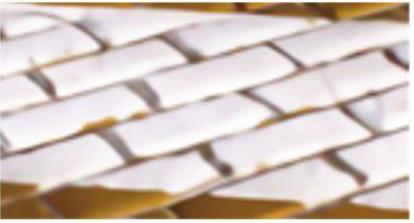}
  \subcaption*{b}
  \end{minipage}
  \begin{minipage}{0.3\textwidth}
  \centering
  \includegraphics[width=1\textwidth]{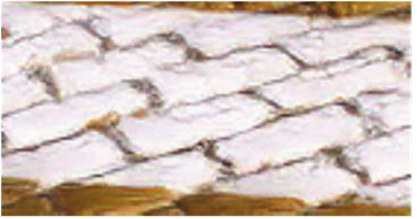}
  \subcaption*{c}
  \end{minipage}
  \end{minipage}

  \begin{minipage}{1\textwidth}
  \centering
  \begin{minipage}{0.3\textwidth}
  \centering
  \includegraphics[width=1\textwidth]{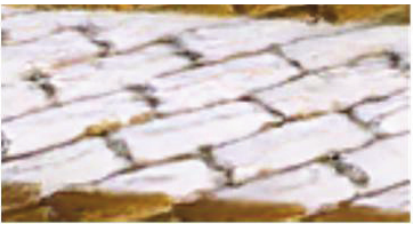}
  \subcaption*{d}
  \end{minipage}
  \begin{minipage}{0.3\textwidth}
  \centering
  \includegraphics[width=1\textwidth]{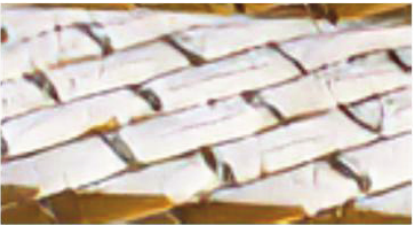}
  \subcaption*{e}
  \end{minipage}
  \begin{minipage}{0.3\textwidth}
  \centering
  \includegraphics[width=1\textwidth]{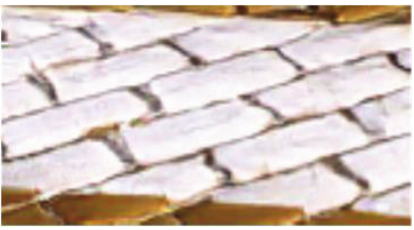}
  \subcaption*{f}
  \end{minipage}
  \end{minipage}

  \begin{minipage}{1\textwidth}
  \centering
  \begin{minipage}{0.3\textwidth}
  \centering
  \includegraphics[width=0.99\textwidth]{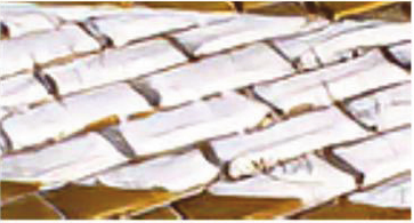}
  \subcaption*{g}
  \end{minipage}
  \begin{minipage}{0.3\textwidth}
  \centering
  \includegraphics[width=1\textwidth]{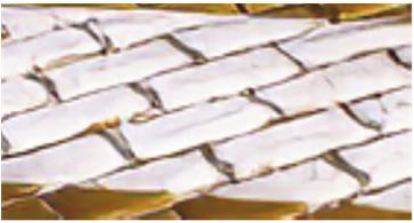}
  \subcaption*{h}
  \end{minipage}
  \begin{minipage}{0.3\textwidth}
  \centering
  \includegraphics[width=1\textwidth]{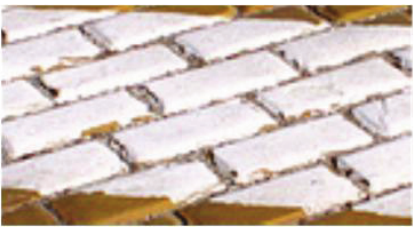}
  \subcaption*{i}
  \end{minipage}
  \end{minipage}
\end{minipage}

\captionsetup{labelfont={bf}}
\caption{Comparison of results of different SR methods on the ``Img091" image from Urban100. (\textbf{a})Bicubic. (\textbf{b})DRN. (\textbf{c})EnhanceNet. (\textbf{d})SRGAN. (\textbf{e})ESRGAN. (\textbf{f})SFTGAN. (\textbf{g})ESRGAN+.  (\textbf{h})Ours. (\textbf{i})Original}
\label{fig:9} 
\end{figure*}

As can be seen from Table \ref{tab:5}, compared with ESRGAN, ESRGAN-DP shows a better performance on SSIM, PSNR and LPIPS. Especially in the four test datasets, LPIPS decreases by 0.01305 on average compared with the ESRGAN, and it is the best among all the models, but SSIM and PSNR are not the best. However, LPIPS is more convincing, because it is more in line with human perceptual habits from a visual perspective. For example, the PSNR value and SSIM value of DRN are mostly the best in all models, but it does not mean that it has the best visual quality. From Fig. \ref{fig:8} and Fig. \ref{fig:9}, too little reasoning about texture details causes the images reconstructed by DRN to still have a certain sense of blur.
\par
We can make a more detailed observation from the fingernails and the edge of sleeves in Fig. \ref{fig:8} and the shape and shadow of bricks in Fig. \ref{fig:9}. These images contain a lot of interactive lines, which contribute to the comparison of the reconstruction ability of SR methods in complex scenes. It is observed that the images reconstructed by Bicubic are extremely blurry in vision. The images reconstructed by EnhanceNet and SRGAN make an improvement in visual quality, but have serious artifacts and distortions. The artifacts of the images reconstructed by ESRGAN are reduced, but too many unreal textures are generated, which results in a great difference between reconstructed images and GT images. SFTGAN performs better than the above methods, but the reconstructed images are poor in the clarity of texture. The images reconstructed by DRN retain more original information, but are still slightly smooth on the whole, which have a certain influence on the visual quality. In terms of clarity of textures, ESRGAN+ has a certain improvement compared with other models, but it also generates too many unrealistic structures to make the images lack facticity. Our proposed method enhances the ability to reason missing information in LR to restore the facticity of the image as much as possible, which makes the reconstructed image more clear and closer to GT image from visual perspective.\par
To sum up, in terms of both evaluation metrics and visual effect, compared with the original perceptual loss only using a single VGG loss, the perceptual loss using DP Loss has superior advantages in solving SR problems.

\begin{table}[htbp]
\captionsetup{labelfont={bf}}
\caption{Comparison of different SR methods on the benchmark datasets}
\label{tab:5}       % Give a unique label
\begin{threeparttable}
\resizebox{\textwidth}{!}{
	\begin{tabular}{llllllllll}
	\toprule[1pt]
	\makecell[l]{DataSet}& \makecell[l]{Metric} & \makecell[l]{Bicubic} & \makecell[l]{DRN\\\citep{37}}& \makecell[l]{EnhanceNet\\\citep{21}} 
     & \makecell[l]{SRGAN\\\citep{17}}& \makecell[l]{ESRGAN\\\citep{22}}& \makecell[l]{SFTGAN\\\citep{23}} & \makecell[l]{ESRGAN+\\\citep{50}} & \makecell[l]{ESRGAN-DP \\(Ours)}\\
	\cmidrule(lr){1-2} 	\cmidrule(lr){3-10}%画线
	
    \multirow{3}*{\shortstack{Set5}}
	& PSNR & 26.69 & 29.95 & 26.76  & 26.69 & 26.50 & 27.26 & 25.88 & 27.11 \\
	& SSIM & 0.7736 & 0.8522 & 0.7670  & 0.7813 & 0.7565 & 0.7765 & 0.7511 & 0.7748 \\
	& LPIPS & 0.3644 & 0.1964 & 0.1198  & 0.1304 & 0.1080 & 0.1028 & 0.1178 & \textbf{0.0990}  \\
    \midrule
    \multirow{3}*{\shortstack{Set14}}
	& PSNR & 26.08 & 28.96 & 26.02  & 25.88 & 25.52 & 26.29 & 25.01 & 26.00 \\
	& SSIM & 0.7466 & 0.8261 & 0.7344  & 0.7347 & 0.7175 & 0.7397 & 0.7159 & 0.7366 \\
	& LPIPS & 0.3870 & 0.2196 & 0.1337  & 0.1422 & 0.1254 & 0.1177 & 0.1363 & \textbf{0.1139}  \\
	\midrule
    \multirow{3}*{\shortstack{BSD100}}
	& PSNR & 26.07 & 25.57 & 25.51  & 24.66 & 24.95 & 25.71 & 24.62 & 25.40 \\
	& SSIM & 0.7177 & 0.7239 & 0.6974  & 0.7063 & 0.6785 & 0.7065 & 0.6893 & 0.6993 \\
	& LPIPS & 0.4454 & 0.2922 & 0.1611  & 0.1622 & 0.1428 & 0.1358 & 0.1446 & \textbf{0.1280}  \\
	\midrule
    \multirow{3}*{\shortstack{Urban100}}
	& PSNR & 24.73 & 26.23 & 24.65  & 24.04 & 24.21 & 25.04 & 23.98 & 24.79 \\
	& SSIM & 0.7101 & 0.7793 & 0.7168  & 0.7209 & 0.7045 & 0.7314 & 0.7182 & 0.7284 \\
	& LPIPS & 0.4346 & 0.2320 & 0.1522  & 0.1534 & 0.1355 & 0.1259 & 0.1334 & \textbf{0.1186} \\
	\bottomrule[1pt]
	\end{tabular}
}
      \begin{tablenotes} %添加此处
		\item \textbf{Font bold} indicates the best performance.
      \end{tablenotes} %添加此处
\end{threeparttable} %添加此处
\end{table}

\subsection{Ablation study}
\label{sec:4.5}
\begin{figure*}[htbp]
\centering
 \begin{minipage}{0.2\textwidth} % change this
 \centering
  \includegraphics[width=1\textwidth]{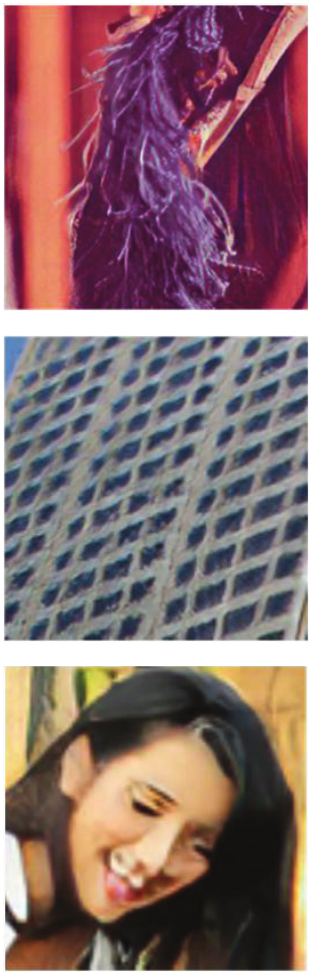}
  \subcaption*{a}
  \end{minipage}
 \begin{minipage}{0.198\textwidth} % change this
 \centering
  \includegraphics[width=1\textwidth]{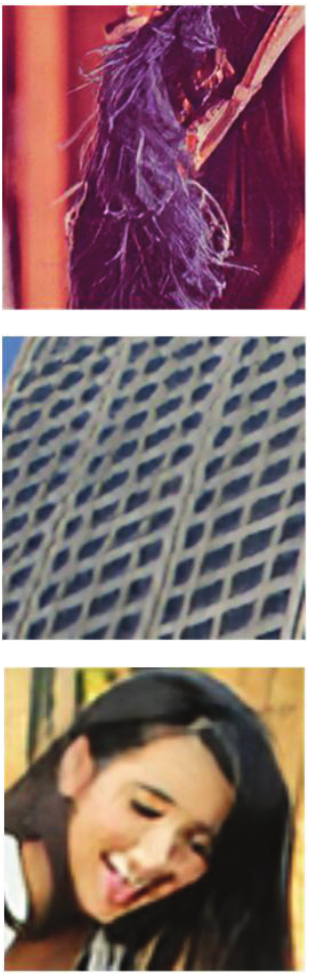}
  \subcaption*{b}
  \end{minipage}
 \begin{minipage}{0.2\textwidth} % change this
 \centering
  \includegraphics[width=1\textwidth]{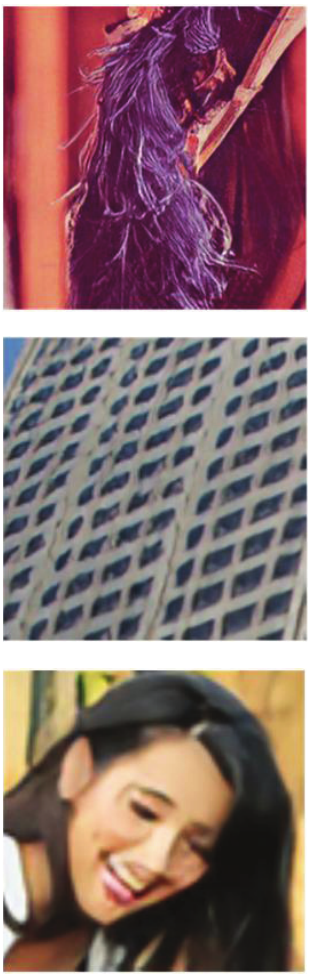}
  \subcaption*{c}
  \end{minipage}
 \begin{minipage}{0.2\textwidth} % change this
 \centering
  \includegraphics[width=1\textwidth]{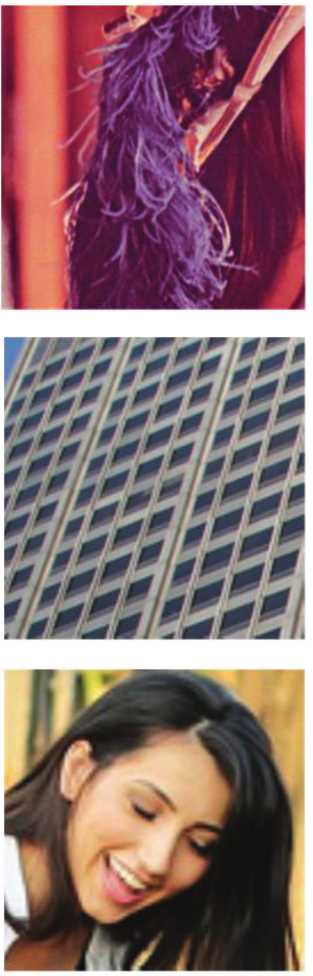}
  \subcaption*{d}
  \end{minipage}

\captionsetup{labelfont={bf}}
\caption{Visual comparison of results of ESRGAN under different conditions. (\textbf{a})VGG loss. (\textbf{b})VGG loss + ResNet loss. (\textbf{c})VGG loss + ResNet loss + Dynamic weighting. (\textbf{d})Original}
\label{fig:10} 
\end{figure*}

In order to demonstrate the effectiveness of the proposed DP Loss and dynamic weighting, ESRGAN is used to as the baseline model to conduct ablation study. It can be seen from Table \ref{tab:6} that whether ResNet loss is applied to the original ESRGAN or the dynamic weighting is applied on the basis of ResNet loss, all evaluation metrics have been improved to a certain extent. Furthermore, from the perspective of SSIM, applying ResNet Loss to ESRGAN can significantly improve the value of SSIM. It shows that the ability of the network to recover the realistic structures of images is greatly improved due to the complementary property of VGG features and ResNet features. From the perspective of LPIPS, both two perceptual losses and dynamic weighting make the images have a great positive impact, which is closely related to the improvement of the visual quality of images. It can be seen from the Fig. \ref{fig:10} that the application of ResNet loss can eliminate a large number of unrealistic artifacts, and then further application of dynamic weighting can make the texture clearer and the structure more in line with GT images.

\begin{table}[htbp]
\captionsetup{labelfont={bf}}
\caption{Comparison of ESRGAN under different conditions}
\label{tab:6}       % Give a unique label
\begin{threeparttable}
	\begin{tabular}{lcccllll}
	\toprule[1pt]
	 \makecell[l]{Metric} & \makecell[c]{VGG \\ loss} & \makecell[c]{ResNet \\ loss}& \makecell[c]{Dynamic \\ weighting} 
     & \makecell[l]{Set5}& \makecell[l]{Set14}& \makecell[l]{BSD100}& \makecell[l]{Urban100}\\
	\cmidrule(lr){1-1} \cmidrule(lr){2-4}	\cmidrule(lr){5-8}%画线
	
    \multirow{3}*{\shortstack{PSNR}}
	& \checkmark &   &   & 26.50 & 25.52 & 24.95 & 24.21 \\
	& \checkmark & \checkmark &  & {\color{blue}26.92} & {\color{blue}25.97} & {\color{blue}25.33} & {\color{blue}24.49}  \\
	& \checkmark & \checkmark & \checkmark & {\color{red}27.11} & {\color{red}26.00} & {\color{red}25.40} & {\color{red}24.79}  \\
    \midrule
    \multirow{3}*{\shortstack{SSIM}}
	& \checkmark &   &   & 0.7565 & 0.7175 & 0.6785 & 0.7045 \\
	& \checkmark & \checkmark &  & {\color{blue}0.7711} & {\color{blue}0.7347} & {\color{blue}0.6945} & {\color{blue}0.7181}  \\
	& \checkmark & \checkmark & \checkmark & {\color{red}0.7748} & {\color{red}0.7366} & {\color{red}0.6993} & {\color{red}0.7284}  \\
	\midrule
    \multirow{3}*{\shortstack{LPIPS}}
	& \checkmark &   &   & 0.1080 & 0.1254 & 0.1428 & 0.1355  \\
	& \checkmark & \checkmark &  & {\color{blue}0.1017} & {\color{blue}0.1165} & {\color{blue}0.1379} & {\color{blue}0.1302}  \\
	& \checkmark & \checkmark & \checkmark & {\color{red}0.0990} & {\color{red}0.1139} & {\color{red}0.1280} & {\color{red}0.1186}  \\
	\bottomrule[1pt]
	\end{tabular}

      \begin{tablenotes} %添加此处
		\item The best performance is highlighted in {\color{red}red } (1st best) and {\color{blue} blue} (2nd best).
      \end{tablenotes} %添加此处
\end{threeparttable} %添加此处
\end{table}

\section{Conclusions}
\label{sec:5}
In this paper, we propose a DP Loss to reslove the problem that there are always structural distortions in the results reconstructed by SR methods under the traditional perceptual-driven. Both the features extracted by pre-trained ResNet network and those extracted by pre-trained VGG network are applied to the perceptual loss, which improves the information acquisition ability of the perceptual features from the perspective of  feature extraction ways to enhance the ability of network to reason texture details. The strategy of using the dynamic weighting method for ResNet loss is to eliminate the interference of the magnitude difference on the training, which makes the advantages of dual perceptual more obvious. In addition, we compare the different influences of different hyperparameter combinations in the DP Loss on the results, and the quantitative and qualitative assessments obtained from four popular benchmark datasets both demonstrate the effectiveness of the method proposed in this paper.



\bibliographystyle{spbasic_unsort}      % basic style, author-year citations
\bibliography{reference}   % name your BibTeX data base

%\begin{thebibliography}{}
% etc
%\end{thebibliography}

\end{document}